First and corresponding author
Konrad Szocik
Department of Social Sciences,
University of Information Technology and Management in Rzeszow
Sucharskiego 2 St., 35-225, Rzeszow, Poland
E-mail: kszocik@wsiz.edu.pl
ORCID: 0000-0002-7262-3915

&

Abraham Loeb
Astronomy Department, Harvard University
60 Garden Street, Cambridge MA 02138, USA
E-mail: aloeb@cfa.harvard.edu

# When Should Science Act on an Interstellar Anomaly?

## Abstract

An unusual interstellar object may warrant further observation even when there is little evidence that it is artificial. The challenge is deciding which observations are worth pursuing, how much to invest in them, and when to stop. We argue that the Loeb Scale can help guide these decisions, provided its levels are understood as provisional assessments rather than estimates of the probability of extraterrestrial technology. Candidate technosignature status may justify targeted follow-up when new observations could distinguish competing explanations and the expected scientific value warrants the cost. This requires testable predictions and retrodictions, attention to natural explanations not yet considered, and recognition of the risks of false alarms. We place this argument in the broader history and philosophy of scientific anomalies and relate it to existing assessment frameworks in SETI and astrobiology. The cases of 1I/‘Oumuamua, 2I/Borisov, and 3I/ATLAS illustrate why classifications and research priorities must remain open to revision. Candidate status is a reason to seek better evidence while it can still be obtained, not a conclusion about an object's origin.

## Introduction

The scientific problem raised by anomalous interstellar objects is often framed too narrowly. Public debate tends to ask whether a given object is natural or artificial, or whether the extraterrestrial hypothesis is respectable enough to be included in scientific discussion. That framing obscures the more important methodological question: when should an unusual object change what scientists do? In other words, when does an anomaly become serious enough to justify more observations, broader comparison of explanations, and a reallocation of scarce attention, without yet justifying belief that the object is technological?

Our focus is on physical interstellar objects identified through their trajectories and observed during their passage through the Solar System. We use "anomalous interstellar object" to describe an object whose measured properties are difficult to explain with available models and data. The argument does not cover all astrophysical anomalies. For example, the gamma-ray signatures of hypothetical antimatter-powered spacecraft discussed by Harris (1986) would require a different observational approach. Unidentified anomalous phenomena (UAPs) also raise distinct evidential and social questions and fall outside the scope of our analysis.

Contemporary work in the philosophy of science and the epistemology of scientific discovery allows us to separate this issue from past tendencies to dismiss it a priori. Lomas argues that extraterrestrial explanations for some anomalous phenomena should not be excluded in advance, especially given the legitimacy of astrobiology, technosignature research, and studies of interstellar travel (Lomas, 2024). Lane similarly contends that the extraterrestrial hypothesis can be a rational explanatory option when the evidence in a particular case makes it relevant, and that the taboo surrounding it often functions as a sociological barrier rather than as an epistemic argument (Lane, 2025). These arguments challenge the current dominant assumption that scientific and academic rigor requires silence about artificial possibilities.

Yet admissibility is not the same as methodological priority. A hypothesis may be scientifically discussable and still too weak, too underdeveloped, or too poorly supported to guide an observational campaign. Cowie emphasizes this point in his analysis of 1I/`Oumuamua. The failure of available natural explanations to satisfy everyone does not by itself support an extraterrestrial-artifact interpretation, because priors, unconceived natural alternatives, and the limited evidential record remain central (Cowie, 2023). His later critique of optimism in the search for extraterrestrial life makes a parallel point at a broader scale: the abundance of planets is not sufficient to establish that life, technological intelligence, or detectable artifacts are common (Cowie, 2025).

The task, then, is not to choose between unrestricted openness and premature belief. It is to decide, case by case, when further observations are likely to reduce an important uncertainty at an acceptable cost. The Loeb Scale helps organize such decisions, but its levels cannot replace scientific judgment or establish a universal threshold for action. By distinguishing ordinary interstellar objects, increasingly anomalous cases, candidate technosignatures, and confirmed technological origin, it creates a vocabulary for escalation before confirmation (Eldadi et al., 2025). Quantitative and evolving versions of the scale make the same point in more operational terms, translating anomaly profiles into scores that can change as data accumulate (Trivedi & Loeb, 2025a, 2025b). The proposed Comprehensive InterStellar Objects Network (CISON) extends the logic institutionally by linking discovery, characterization, prioritization, and possible interception (Trivedi & Loeb, 2026). In this article, we explain how candidate status can justify a change in scientific practice without implying belief in an artificial origin. We show that the status of technosignature candidate constitutes an intermediate epistemic category. It is stronger than merely allowing for the possibility of proposing a hypothesis about the existence of extraterrestrial technology, weaker than any claim of confirmation, and yet capable of guiding targeted observations when competing explanations can be tested. We suggest that this change signals to the scientific community that the matter in question warrants more intensive evidence-gathering, but it does not tell the community what the object in question is.

**From Possibility to Research Priority**

The debate after Lomas, Lane, and Cowie can be understood as a dispute about the transition from permissibility to priority. Lomas and Lane are persuasive against a form of skeptical gatekeeping that treats the extraterrestrial hypothesis as disqualified before case-specific analysis begins. The neighboring sciences of exoplanet research, astrobiology, SETI, technosignature studies, and propulsion physics make such blanket exclusion difficult to defend (Lomas, 2024; Lane, 2025). Artificial-origin hypotheses need not violate established science, and they need not be treated as conceptually incoherent simply because they are extraordinary in cultural imagination.

However, removing a taboo does not provide a decision rule. Scientific inquiry must still determine when an anomaly is strong enough to affect telescope time, follow-up priorities, data-sharing obligations, and the range of hypotheses that receive systematic development. A field can be too closed, but it can also be too easily redirected by dramatic but poorly constrained anomalies. Openness is therefore a necessary background virtue. Because reputational bias is rarely acknowledged explicitly, the practical response is to assess research proposals against transparent criteria, including data quality, testable predictions, and expected scientific value, rather than the reputation of the hypothesis or its proponents. Cowie's critique marks the opposite boundary. In the case of 1I/`Oumuamua, the relevant evidence included unusual shape or light curve inferences, non-gravitational acceleration, and the absence of clear cometary activity in some observations (Meech et al., 2017; Micheli et al., 2018). These features made the object unusually interesting, and they motivated proposals such as the solar-radiation-pressure hypothesis for a very thin object (Bialy & Loeb, 2018; Loeb, 2022). But Cowie is right that an incomplete natural explanation does not automatically raise an artificial explanation to high probability (Cowie, 2023). A weak natural account can coexist with a still weaker artificial one, and unknown natural categories may later become visible.

The needed category lies somewhere between. A candidate technosignature is not a conclusion that an object is technological. It is a conclusion that the case has crossed a threshold at which ordinary handling is no longer adequate. The object should be investigated with greater care because the cost of missing a significant signal may be high and because the window for gathering data from interstellar visitors is often short. This is especially important for objects passing rapidly through the inner Solar System, where delayed recognition can permanently foreclose decisive observations.

The distinction between belief and escalation is crucial. Scientific communities routinely act under uncertainty without assigning high credence to a favored explanation. They run follow-up experiments, allocate observing time, and build classifications because early evidence is ambiguous. For example, follow-up spectroscopy of 2I/Borisov revealed its unusually high carbon monoxide abundance, a scientifically valuable finding independent of any technological hypothesis (Bodewits et al., 2020). A candidate technosignature should be treated in the same way, as a reason to improve the evidential situation.

**The Loeb Scale and Research Decisions**

The Loeb Scale converts a vague controversy into a structured problem of scientific response. The first formulation of the scale classifies interstellar objects across levels ranging from ordinary natural objects to confirmed artificial origin, with intermediate levels representing increasing anomaly and increasing relevance of technosignature indicators (Eldadi et al., 2025). Level 4 is especially important because it is the point at which technosignature considerations become formally salient while confirmation remains absent.

Without an intermediate category, anomalous cases tend to be forced into a misleading binary. Either they are treated as ordinary natural objects until artificial origin is proven, or they are discussed as if anomaly already licenses dramatic conclusions. Both options distort inquiry. The first can suppress attention before adequate evidence is collected. The second can inflate weak evidence into premature claims. Level 4 creates a third option: organized escalation under explicit uncertainty.

Graded assessment before confirmation is not new. The Rio Scale and Rio 2.0 address the significance and communication of possible SETI detections (Almár & Tarter, 2011; Forgan et al., 2019). Green et al. (2021) propose a framework for reporting evidence of life beyond Earth, while Vickers et al. (2023) show, in the context of biosignatures, how unconceived alternatives limit confidence in detection claims. We examine how the provisional classification of an interstellar object should affect decisions about further observations. The Loeb Scale can help organize this decision, but its scores do not measure the probability of artificial origin. Their interpretation depends on the available data, competing models, and background assumptions. The quantitative mapping of the scale makes this intermediate status more tractable. Trivedi and Loeb translate categorical levels into a continuous framework using observable metrics such as, among others, non-gravitational acceleration, unusual geometry, trajectory improbability, and possible operational behavior (Trivedi & Loeb, 2025b). The value of this model lies in the fact that it clearly identifies which characteristics raise concerns and to what extent they intersect across different dimensions.

Loeb's evolving scale introduces a temporal dimension. Interstellar objects are rarely fully understood at the time of their initial detection. Their assessments should therefore be able to change as new observations come in, while retaining information about earlier evidence and avoiding unstable reactions to short-term fluctuations (Trivedi & Loeb, 2025a). Candidate status is not a one-time label. It is a dynamic assessment of whether a given case continues to warrant special attention as the body of evidence evolves.

CISON turns the same logic into an institutional design. A network that links wide-field discovery, rapid high-resolution characterization, and selective mission escalation would make anomaly assessment proactive rather than merely retrospective (Trivedi & Loeb, 2026). Such a system would be valuable even if every anomalous object ultimately received a natural explanation. This would improve the discipline's ability to secure fleeting evidence, systematically compare cases, and avoid decisions motivated primarily by prestige, publicity, or concern for reputation.

## Anomalies in the History of Science

Historical cases do not prove that any present anomaly is a technosignature. They show that communities can fail to respond properly to serious evidence when theory, institutional incentives, or professional identity make the evidence difficult to assimilate.

The philosophical problem of scientific anomalies has a long history. Kuhn (1962) examined their role in established scientific traditions, while Laudan (1977) understood scientific progress in terms of solving empirical and conceptual problems. The possibility of extraterrestrial intelligence was discussed by Beck (1971) and Lem (2013, Chapter 3), while Ćirković (2018, 2023) examined methodological problems in identifying possible extraterrestrial artifacts. For interstellar objects, the practical issue is which further observations could distinguish competing explanations and whether they should be prioritized.Kaplan's discussion of puerperal fever illustrates the point sharply. Semmelweis identified a stable and consequential pattern before modern germ theory was available: mortality differed dramatically between wards, and the difference was linked to physicians moving from dissections to childbirth without effective handwashing. When handwashing was introduced, mortality declined (Kaplan, 2026). The evidential pattern was serious before the explanatory framework was complete. Resistance arose not only from uncertainty, but also from the implication that physicians themselves were causing preventable deaths. Gordon and Holmes, as Kaplan presents them, show that convergence across cases can provide warrant even before a single decisive theoretical synthesis has been accepted (Kaplan, 2026).

Pierre Louis adds a methodological lesson to this. His statistical comparisons challenged established medical practices, such as bloodletting, showing that trust based on habit and professional authority could persist despite the lack of evidence for the effectiveness of these methods (Kaplan, 2026). The analogy to anomaly assessment is that tacit confidence can resist discipline by explicit comparison. A strange interstellar object may be normalized too quickly if the community lacks procedures for comparing anomalies across multiple dimensions.

These cases support our modest conclusion that serious anomalies require protection from both dismissal and overinterpretation. They need institutional space, comparative methods, and enough time for evidence to mature. The status of potential candidate technosignature should serve precisely this function and support further scientific inquiry, rather than favoring one explanation over another. These examples do not show that formal institutional support is necessary for discovery. They suggest only that clear procedures may help researchers obtain important evidence before it is lost.

## What Candidate Status Requires

A candidate technosignature is best defined in the present context as a physical interstellar object whose observed properties justify targeted tests of competing natural and, where testable, technological explanations, while leaving the question of origin unresolved. This definition has five components. First, candidate status must be data-dependent. The Galileo Project emphasizes transparent, high-quality, multimodal data as a condition for distinguishing natural, human-made, and anomalous phenomena (Loeb & Laukien, 2023). Watters et al. similarly describe a framework in which outliers are identified within a high-dimensional observational space constructed from multiple sensor modalities (Watters et al., 2023). The same standard of data quality, though not necessarily the same observational methods, should apply to interstellar objects. UAPs and interstellar objects differ in their observational conditions, possible sources of error, and social context. We refer to UAP research only to illustrate general standards of evidence, not to suggest that both categories require the same assessment framework. Second, candidate status is cumulative. No single irregularity should normally carry the whole burden. The evidential profile becomes more serious when independent dimensions converge: trajectory, acceleration, spectrum, shape, surface properties, electromagnetic features, and behavior relative to solar radiation or gravitational expectations (Eldadi et al., 2025; Trivedi & Loeb, 2025b). This cumulative structure also connects with broader technosignature assessment frameworks, which emphasize graded evaluation rather than simple yes-or-no classification (Lingam et al., 2023).

Third, candidate status remains compatible with underdetermination. Unconceived natural explanations remain possible, and in many cases likely. Cowie's warning is therefore a constraint on its interpretation (Cowie, 2023, 2025). The point of the category is to prevent the explanatory field from closing too early in the opposite direction.

Fourth, candidate status is action-guiding. Formal consideration means identifying which observations could distinguish competing explanations and deciding whether these observations are worth pursuing. Investigators should specify the main anomaly, the leading natural explanations, and any testable technological hypothesis. They should examine both predictions for future observations and retrodictions concerning existing data. The aim is to test competing models, not merely to adjust them after each new observation. Once a case crosses the relevant threshold, it may become rational to intensify observation, preserve and share data, broaden modeling efforts, and, in exceptional circumstances, consider mission-level escalation (Eldadi et al., 2025; Trivedi & Loeb, 2025a, 2025b, 2026). These decisions must account for expected information gain, observing time, financial costs, and competing research priorities. A high classification score alone cannot justify costly escalation. This is the primary epistemic significance of Level 4, which is intended to enable a change in practice without, however, justifying the belief that a given object is artificial. Fifth, candidate status is revisable. New data may reduce the score, strengthen a natural explanation, reveal an instrumental artifact, or show that the anomaly was transient. Conversely, new evidence may increase the score and justify further escalation. Special priority should be withdrawn when available natural explanations adequately account for the observations or further investigation is unlikely to produce useful evidence. Public communication must also distinguish a provisional classification from the probability of artificial origin. Presenting a score as an estimate of "alien likelihood" risks false alarms, unnecessary public concern, and damage to the credibility of technosignature research (Forgan et al., 2019; Green et al., 2021).

## Comparing Three Interstellar Objects

1I/`Oumuamua remains valuable precisely because it is unresolved and contested. It was the first recognized interstellar object, and its observational profile included features that made ordinary classification difficult (Meech et al., 2017; Micheli et al., 2018). Bialy and Loeb proposed that solar radiation pressure acting on an unusually thin object could explain its peculiar acceleration, and Loeb later developed the possibility of artificial origin as a hypothesis that should not be excluded in advance (Bialy & Loeb, 2018; Loeb, 2022). Critics, including Cowie, have argued that the evidential basis is insufficient for an artifact conclusion, especially given uncertainty about priors and possible unknown natural alternatives (Cowie, 2023).

This disagreement is exactly why the candidate category is needed. Treating 1I/`Oumuamua as a candidate does not settle its origin. It says that the object was anomalous enough that a more organized response would have been scientifically rational if available at the time. The conclusion is that the scientific community needs procedures that allow it to distinguish between "unconfirmed" and "not worth further consideration."

Under the Loeb Scale, a Level 4 interpretation of 1I/'Oumuamua would mean that technosignature indicators deserve a documented plan for further testing, not that the extraterrestrial-artifact hypothesis has won (Eldadi et al., 2025; Trivedi & Loeb, 2025b). Such a classification would have encouraged rapid follow-up, comparison with future interstellar objects, and careful preservation of the anomaly profile. It should be communicated as a provisional observational priority, not as an estimate of the probability of artificial origin. The other two confirmed interstellar objects provide useful comparisons. 2I/Borisov displayed cometary activity and an unusually carbon monoxide-rich coma (Bodewits et al., 2020). Observations of 3I/ATLAS have also established cometary activity and revealed an unusual isotopic composition consistent with formation in a cold and distant environment (Bolin et al., 2025; Cordiner et al., 2026). Eldadi et al. (2025) assigned Level 0 to 2I/Borisov and Level 4 to 3I/ATLAS. These classifications belong to the proposed Loeb Scale and should not be interpreted as probabilities of artificial origin. They illustrate why any observational priority must remain open to revision as natural explanations develop.

**AI in Anomaly Assessment**

Future research and observatories may generate streams of anomalies too extensive to be evaluated solely through human assessment. Artificial intelligence should therefore play a supporting role in the assessment of potential technological signatures, but its role must be carefully limited. It should be used to detect outliers, compare cases, estimate similarity to known classes, prioritize further actions, and clearly justify those priorities.

This role is already visible in current research programs. The Galileo Project includes AI-based analysis within a broader commitment to open, multimodal, reproducible data (Loeb & Laukien, 2023). Watters et al. describe anomaly detection as a problem of identifying unusual events in high-dimensional sensor space (Watters et al., 2023). CISON extends this approach to interstellar objects by connecting discovery and characterization with evolving Loeb Scale classification (Trivedi & Loeb, 2026).

Quantitative methods do not eliminate judgment, but they introduce an element of discipline. When used properly, artificial intelligence can limit selective attention, bias, or overreliance on familiar categories. It can also curb sensationalism by showing which features made a given case unusual and how that case relates to previous natural, artificial, and unsolved cases. In this sense, artificial intelligence is valuable because it structures and organizes our attention. However, AI systems may reproduce biases or misclassify unfamiliar phenomena. Their assessments should remain transparent, independently verifiable, and subject to human oversight. AI should not independently determine whether an anomaly is technological or whether its interpretation should be publicly announced.

**Conclusion**

The main issue in research on anomalous interstellar objects is whether the scientific community has reliable procedures for determining when uncertainty becomes significant enough to warrant taking action. The Loeb Scale, particularly Level 4, and its quantitative and evolving extensions provide a framework for formalizing this situation (Eldadi et al., 2025; Trivedi and Loeb, 2025a, 2025b). CISON demonstrates how such a classification could be integrated into a broader observational architecture (Trivedi and Loeb, 2026). From a philosophical standpoint, candidate status allows for the separation of escalation from endorsement. Hypotheses regarding extraterrestrial technologies should not be ruled out solely because of taboos or reputational concerns (Lomas, 2024; Lane, 2025). At the same time, an anomaly does not constitute evidence strong enough to take an artificial origin for granted, and unforeseen natural explanations must remain viable possibilities (Cowie, 2023, 2025). Candidate status occupies the space in between. Historical examples discussed by Kaplan show that discovery often depends on the anomaly persisting long enough for better theoretical frameworks, instruments, and comparative possibilities to emerge (Kaplan, 2026). A scientifically sound response to anomalous interstellar objects should therefore be neither overly credulous nor dismissive. It should be prepared to devote attention to the matter before certainty is achieved, since in some cases the opportunity to obtain conclusive evidence will be fleeting. We should strive to organize our research in such a way that, should an extraterrestrial technosignature ever come within range of observation, science would be able to detect it in time.

**References**

Almár, I., & Tarter, J. (2011). The discovery of ETI as a high-consequence, low-probability event. Acta Astronautica, 68(3–4), 358–361. https://doi.org/10.1016/j.actaastro.2009.07.007

Beck, L. W. (1971). Extraterrestrial intelligent life. Proceedings and Addresses of the American Philosophical Association, 45, 5–21. https://doi.org/10.2307/3129745

Bialy, S., & Loeb, A. (2018). Could solar radiation pressure explain `Oumuamua's peculiar acceleration? The Astrophysical Journal Letters, 868(1), L1. https://doi.org/10.3847/2041-8213/aaeda8

Bodewits, D., Noonan, J. W., Feldman, P. D., Bannister, M. T., Farnocchia, D., Harris, W. M., Li, J.-Y., Mandt, K. E., Parker, J. Wm., & Xing, Z.-X. (2020). The carbon monoxide-rich interstellar comet 2I/Borisov. Nature Astronomy, 4, 867–871. https://doi.org/10.1038/s41550-020-1095-2

Bolin, B. T., Belyakov, M., Fremling, C., Graham, M. J., Abdelaziz, A. M., Elhosseiny, E., Gray, C. L., Ingebretsen, C., Jewett, G., Lisse, C. M., Karpov, S., Kilic, M., Mašek, M., Molham, M., Roderick, D., Takey, A., Abron, L.-M., Coughlin, M. W., Hsieh, C.-H., Noll, K. S., & Wong, I. (2025). Interstellar comet 3I/ATLAS: Discovery and physical description. Monthly Notices of the Royal Astronomical Society: Letters, 542(1), L139–L143. https://doi.org/10.1093/mnrasl/slaf078

Ćirković, M. M. (2018). Astroengineering, Dysonian SETI, and naturalism: A new Catch-22. Acta Astronautica, 152, 289–298. https://doi.org/10.1016/j.actaastro.2018.07.051

Ćirković, M. M. (2023). Alien technology, conjunction and ergodicity. International Journal of Astrobiology, 22(3), 197–204. https://doi.org/10.1017/S147355042200043X

Cordiner, M., Roth, N. X., Micheli, M., Villanueva, G., Farnocchia, D., Charnley, S., Biver, N., Bockelée-Morvan, D., Bodewits, D., Chandler, C. O., Crovisier, J., Drozdovskaya, M. N., Furuya, K., Kelley, M. S. P., Milam, S., Noonan, J. W., Opitom, C., Schwamb, M. E., & Thomas, C. A. (2026). Isotopic evidence for a cold and distant origin of 3I/ATLAS. Nature, 655, 870–874. https://doi.org/10.1038/s41586-026-10771-6

Cowie, C. (2023). Arguing about extraterrestrial intelligence. The Philosophical Quarterly, 73(1), 64-83. https://doi.org/10.1093/pq/pqac009

Cowie, C. (2025). Optimism in the search for extraterrestrial life? A philosophical perspective. Proceedings of the Aristotelian Society, 125(1), 44-61. https://doi.org/10.1093/arisoc/aoaf003

Eldadi, O., Tenenbaum, G., & Loeb, A. (2025). The interstellar object significance scale (Loeb Scale): Astronomical classification of interstellar objects. International Journal of Astrobiology, 24, e22. https://doi.org/10.1017/S1473550425100190

Forgan, D., Wright, J., Tarter, J., Korpela, E., Siemion, A., Almár, I., & Piotelat, E. (2019). Rio 2.0: Revising the Rio scale for SETI detections. International Journal of Astrobiology, 18(4), 336–344. https://doi.org/10.1017/S1473550418000162

Green, J., Hoehler, T., Neveu, M., Domagal-Goldman, S., Scalice, D., & Voytek, M. (2021). Call for a framework for reporting evidence for life beyond Earth. Nature, 598, 575–579. https://doi.org/10.1038/s41586-021-03804-9

Harris, M. J. (1986). On the detectability of antimatter propulsion spacecraft. Astrophysics and Space Science, 123, 297–303. https://doi.org/10.1007/BF00653949

Kaplan, M. (2026). I Told You So!: Scientists Who Were Ridiculed, Exiled, and Imprisoned for Being Right. St. Martin's Griffin.

Kuhn, T. S. (1962). The structure of scientific revolutions. University of Chicago Press.

Lane, W. C. (2025). The extraterrestrial hypothesis: An epistemological case for removing the taboo. European Journal for Philosophy of Science, 15, Article 8. https://doi.org/10.1007/s13194-025-00634-8

Laudan, L. (1977). Progress and its problems: Towards a theory of scientific growth. Routledge & Kegan Paul.

Lem, S. (2013). Summa technologiae (J. Zylinska, Trans.). University of Minnesota Press. (Original work published 1964).

Lingam, M., Haqq-Misra, J., Wright, J. T., Huston, M. J., Frank, A., & Kopparapu, R. (2023). Technosignatures: Frameworks for their assessment. The Astrophysical Journal, 943(1), Article 27. https://doi.org/10.3847/1538-4357/acaca0

Loeb, A. (2022). On the possibility of an artificial origin for `Oumuamua. Astrobiology, 22(12), 1392-1399. https://doi.org/10.1089/ast.2021.0193

Loeb, A., & Laukien, F. H. (2023). Overview of the Galileo Project. Journal of Astronomical Instrumentation, 12(1), 2340003. https://doi.org/10.1142/S2251171723400032

Lomas, T. (2024). The extraterrestrial hypothesis: A case for scientific openness to an interstellar explanation for unidentified anomalous phenomena. Philosophy and Cosmology, 32, 34-59. https://doi.org/10.29202/phil-cosm/32/3

Meech, K. J., Weryk, R., Micheli, M., Kleyna, J. T., Hainaut, O. R., Jedicke, R., Wainscoat, R. J., Chambers, K. C., Keane, J. V., Petric, A., Denneau, L., Magnier, E. A., Berger, T., Huber, M. E., Flewelling, H. A., Waters, C., Schunova-Lilly, E., & Chastel, S. (2017). A brief visit from a red and extremely elongated interstellar asteroid. Nature, 552, 378-381. https://doi.org/10.1038/nature25020

Micheli, M., Farnocchia, D., Meech, K. J., Buie, M. W., Hainaut, O. R., Prialnik, D., Schorghofer, N., Weaver, H. A., Chodas, P. W., Kleyna, J. T., Weryk, R., Wainscoat, R. J., Ebeling, H., Keane, J. V., Chambers, C. K., Koschny, D., & Petropoulos, A. E.

(2018). Non-gravitational acceleration in the trajectory of 1I/2017 U1 (`Oumuamua). Nature, 559, 223-226. https://doi.org/10.1038/s41586-018-0254-4

Trivedi, O., & Loeb, A. (2025a). Evolving the Loeb Scale [Preprint]. arXiv. https://arxiv.org/abs/2512.13743

Trivedi, O., & Loeb, A. (2025b). Quantitative mapping of the Loeb Scale [Preprint]. arXiv. https://arxiv.org/abs/2509.06253

Trivedi, O., & Loeb, A. (2026). A comprehensive network for the discovery and characterization of interstellar objects [Preprint]. arXiv. https://arxiv.org/abs/2601.21184

Vickers, P., Cowie, C., Dick, S. J., Gillen, C., Jeancolas, C., Rothschild, L. J., & McMahon, S. (2023). Confidence of life detection: The problem of unconceived alternatives. Astrobiology, 23(11), 1202–1212. https://doi.org/10.1089/ast.2022.0084

Watters, W. A., Loeb, A., Laukien, F., Cloete, R., Delacroix, A., Dobroshinsky, S., Horvath, B., Kelderman, E., Little, S., Masson, E., Mead, A., Randall, M., Schultz, F., Szenher, M., Vervelidou, F., White, A., Ahlstrom, A., Cleland, C., Dockal, S., Donahue, N., Elowitz, M., Ezell, C., Gersznowicz, A., Gold, N., Hercz, M. G., Keto, E. R., Knuth, K. H., Lux, A., Melnick, G. J., Moro-Martin, A., Martin-Torres, J., Llusa Ribes, D., Sail, P., Teodorani, M., Tedesco, J. J., Tedesco, G. T., Tu, M., & Zorzano, M.-P. (2023). The scientific investigation of unidentified aerial phenomena (UAP) using multimodal ground-based observatories. Journal of Astronomical Instrumentation, 12(1), 2340006. https://doi.org/10.1142/S2251171723400068